\documentstyle[12pt]{article}
\begin{document}
\begin{center}
{\large\bf The Exact Solutions of Some Multidimensional
Generalizations of the Fokker-Planck 
Equation }

{\large\bf used by R.Friedrich and J.Peinke for the Description}

{\large\bf  of a Turbulent Cascade}

\vspace*{0.15truein}
{A. A. Donkov$^1$, A. D. Donkov$^2$ and E. I. Grancharova$^3$}
\vspace*{0.15truein}

$^1${\it Dept.of Physics, University of Wisconsin, 1150 Univ-Ave,
Madison, WI-53706, USA}

e-mail : donkov@phys-next1.physics.wisc.

$^{2}${\it Dept.of Physics, University of Sofia,
 5 J.Bourchier Blvd., Sofia 1164, BULGARIA}

e-mail: donkov@phys.uni-sofia.bg \\
and {\it Bogoliubov Lab. of Theoretical Physics, JINR, Dubna,
Moscow Region, RUSSIA}\\
e-mail: donkov@thsun1.jinr.ru

$^{3}${\it Dept.of Physics, University of Sofia,
 5 J.Bourchier Blvd., Sofia 1164, BULGARIA}\\

e-mail: granch@phys.uni-sofia.bg
\end{center}
\vspace*{0.15truein}

\begin{abstract}
Some multidimensional generalizations of the
Fokker-Planck equation
used by R. Friedrich and J. Peinke for the description of a turbulent
cascade  as a stochastic process of Markovian type,
are considered. The exact solutions of the Cauchy problems for these
equations are found with the operator methods.
\end{abstract}

\section{Introduction}
The understanding of the turbulence is one of the main unsolved problems
of classical physics, in spite of the more than 250 years of
strong investigations initiated by D.Bernoulli and L.Euler.

In the stochastic approach to turbulence~\cite{Monin},~\cite{Frish}
the turbulent cascade is
considered as a stochastic process, described by the probability
distribution $P(\lambda,v)$, where $\lambda$ and $v$ are the
appropriate scaled length and the velocity increment respectively.
Recently~\cite{Friedrich} R.Friedrich and J.Peinke presented
experimental evidence that 
the probability
density function  $P(\lambda,v)$ obeys a Fokker-Planck
equation (FPE)~\cite{Risken} (see fig.1 and fig.2
in~\cite{Friedrich}):
\begin{equation}
\frac{\partial P(\lambda,v)}{\partial \lambda} =
\left[ -\frac{\partial}{\partial v} D^1(\lambda,v)
+ \frac{\partial^2}{\partial v^2} D^2(\lambda,v) \right]
P(\lambda,v),
\label{FPP}
\end{equation}
where the drift and duffusion coefficients
$D^1$ and $D^2$ respectively are derived by analysis of experimental
data of a fluid dynamical experiment (see fig.3 in~\cite{Friedrich}).

In their paper Friedrich and Peinke 
use the following approximations for the drift and diffusion terms
respectively
$$
D^1= -a\, v, \qquad a>0; \qquad
D^2 = c\, v^2,\qquad c>0 \,,
$$
where $a$ and $c$ do not depend on $\lambda$.

In our previous paper~\cite{Donkov1}, using the method of
M.Suzuki~\cite{Suzuki} and the Feynman's disentangling
techniques~\cite{Feynman} we gave the exact solution of the Cauchy problem
for the Eq.~(\ref{FPP}) with more realistic approximations
(see fig.3 in~\cite{Friedrich})
for $D^1$ and $D^2$ :
\begin{equation}
D^1= -a(\lambda)\, v, \qquad a(\lambda)>0 ;\qquad
D^2 = c(\lambda)\, v^2,\qquad c(\lambda)>0 .
\label{D}
\end{equation}

In Section 2 of this paper we consider the Cauchy problem  for the following
n-dimensional generalization of the Eqs.~(\ref{FPP}),~(\ref{D}):
\begin{equation}
\frac{\partial P}{\partial \lambda} = b_0(\lambda)P(\lambda,{\bf v})+
b_1(\lambda) {\bf v}\cdot\nabla_{\bf v} P(\lambda,{\bf v}) +
c(\lambda){\bf v}^2\Delta_{\bf v} P(\lambda,{\bf v}),
\label{P}
\end{equation}
$$
P(0, {\bf v}) = \varphi ({\bf v}).
$$
This equation is derived from the n-dimensional FPE
\begin{equation}
\frac{\partial P}{\partial \lambda} =
-\nabla_{\bf v}\cdot {\bf D}^1(\lambda,{\bf v}) P(\lambda,{\bf v}) +
\sum_{i,j=1}^{n}\frac{\partial^2}{\partial v_i \partial v_j}
D_{ij}^2(\lambda,{\bf v}) P(\lambda,{\bf v})
\label {FP1}
\end{equation}
with the drift vector ${\bf D}^1(\lambda,{\bf v}) = -a(\lambda){\bf v}$
and the diffusion tensor
$\hat D^2(\lambda,{\bf v}) = c(\lambda){\bf v}^2\hat1.$
 (  Consequently
$b_0(\lambda)= n[a(\lambda)+2c(\lambda)]$ and
$b_1(\lambda)=a(\lambda) + 4 c(\lambda)$ ).

In Section 3 we find the exact solution of the Cauchy problem
\begin{equation}
\frac{\partial P}{\partial \lambda} = b_0(\lambda)P(\lambda,{\bf v})+
b_1(\lambda) {\bf v}\cdot\nabla_{\bf v} P(\lambda,{\bf v}) +
c(\lambda)({\bf v}\cdot\nabla_{\bf v})^2  P(\lambda,{\bf v}),
\label{PP}
\end{equation}
$$
P(0, {\bf v}) = \varphi ({\bf v}).
$$
The Eq.~(\ref{PP}) is derived from the FPE~(\ref{FP1})
with the drift vector ${\bf D}^1(\lambda,{\bf v}) = -a(\lambda){\bf v}$
and the diffusion tensor
$\hat D^2(\lambda,{\bf v}) = c(\lambda){\bf v}{\bf v}.$
 ( Consequently
$b_0(\lambda)= na(\lambda)+(n^2+n)c(\lambda)$ and
$b_1(\lambda)=a(\lambda) + (2n+1) c(\lambda)$ ).

Because of the analytic expressions of the diffusion tensor, one may regard
the cases (\ref{P}) and (\ref{PP}) respectively as the "isotropic"
and "degenerate anisotropic" $(\det \hat D =0)$ ones.
\begin{center}
\section{Exact Solution of the Cauchy Problem (\ref{P})}
\end{center}
In the spirit of the operational methods
( see \cite{Dubinskii}$-\!$\cite{Donkov} )
we have for the solution of the problem~(\ref{P})
\begin{equation}
P(\lambda,{\bf v}) =
\left(exp_+ \int_0^{\lambda} \left[ b_0(s)+b_1(s){\bf v}\cdot\nabla_{\bf v}
+c(s){\bf v}^2\Delta_{\bf v}\right]{\rm d}s\right) \varphi({\bf v}),
\label{form}
\end{equation}
where the symbol $\;\;exp_+\int_0^{\lambda}\hat C(s){\rm d}s\;\;$
designates the V.Volterra ordered exponential
\begin{equation}
exp_+ \int_0^{\lambda} \hat C(s) {\rm d}s =
\hat 1 + \lim_{n\to\infty} \sum_{k=1}^n\int_0^{\lambda}{\rm d}\lambda_1
\int_0^{\lambda_1}{\rm d}\lambda_2 \dots \int_0^{\lambda_{k-1}}
{\rm d}\lambda_{k}
\hat C(\lambda_1) \hat C(\lambda_2) \dots \hat C(\lambda_{k}).
\label{exp}
\end{equation}

The linearity of the integral and the explicit form of the operators
in~(\ref{form}) permit to write the solution $P(\lambda,{\bf v})$ in terms
of the usual, not ordered, operator valued exponent
\begin{equation}
P(\lambda,{\bf v}) = {\rm e}^{\beta_0 (\lambda)}\,
{\rm e}^{\beta_1(\lambda){\bf v}.\nabla_{\bf v}
+\gamma (\lambda){\bf v}^2\Delta_{\bf v}}
\varphi({\bf v}) ,
\label{expP}
\end{equation}
where for convenience we have denoted
\begin{equation}
\beta_j(\lambda) = \int_0^{\lambda}b_j(s){\rm d}s,\;\; (j=0,1);
\qquad \gamma(\lambda) = \int_0^{\lambda}c(s) {\rm d}s .
\label{beta}
\end{equation}
Consequently (from now on "$'$" means $\frac{\rm d}{{\rm d}t} $ ) :
\begin{equation}
\beta_j(0)=0,\;\;\; {\beta}'_j(\lambda) =b_j(\lambda),\;\;\; (j=0,1);
\qquad \gamma(0)=0,\;\;\; {\gamma}'(\lambda)=c(\lambda).
\label{betaPR}
\end{equation}

Since
\begin{equation}
\left[\; \beta_1(\lambda) {\bf v}\cdot \nabla_{\bf v}\; ,\;
\gamma(\lambda) {\bf v}^2\Delta_{\bf v}\;\right] = 0
\label{com}
\end{equation}
the solution (\ref{expP}) can be written in the form:
\begin{equation}
P(\lambda, {\bf v})=
{\rm e}^{\beta_0(\lambda)}
{\rm e}^{\beta_1(\lambda) {\bf v}\cdot\nabla_{\bf v}}
{\rm e}^{\gamma(\lambda){\bf v}^2 \Delta_{\bf v}}
\varphi({\bf v})=
{\rm e}^{\beta_0(\lambda)}
{\rm e}^{\gamma(\lambda){\bf v}^2 \Delta_{\bf v}}
{\rm e}^{\beta_1(\lambda) {\bf v}\cdot\nabla_{\bf v}}
\varphi({\bf v})
\label{Form}
\end{equation}

To write the expression (\ref{Form} ) in a final form
we will use the following formulae for acting 
of the pseudodifferential operators~\cite{Hoermander}~-~\cite{Treves}
${\rm e}^{\beta {\bf v}\cdot\nabla_{\bf v}}$ and
${\rm e}^{\tau {\bf v}^2\Delta_{\bf v}}$
on arbitrary functions of {\bf v}:
\begin{equation}
{\rm e}^{\beta(\lambda){\bf v}\cdot \nabla_{\bf v}
}f({\bf v})
= f\left({\bf v}{\rm e}^{\beta(\lambda)}\right)
\label{F1}
\end{equation}
and
 (for $\tau(\lambda) >0$)
\begin{equation}
{\rm e}^{\tau(\lambda){\bf v}^2\Delta_{\bf v}}g({\bf v}) =
\frac{{\rm e}^{\tau(\lambda)\hat\Lambda_{\theta}}}
{\sqrt{4\pi\tau(\lambda)}}
\int_{-\infty}^{\infty}
{\rm e}^{-\frac{s^2}{4\tau(\lambda)}}
g\left({\bf v}{\rm e}^{(n-2)\tau(\lambda)\pm s}\right){\rm d}s.
\label{F2}
\end{equation}
Here $\hat\Lambda_{\theta}$ is the operator of Laplace-Beltrami
on the sphere $S_1$ in ${\cal R}^n$ ~\cite{Bateman}$\!-\!$\cite{Mich}:
\begin{equation}
\Delta_{\bf v} = \frac{\partial^2}{\partial v^2} +
\frac{n-1}{v} \frac{\partial}{\partial v} +
\frac{1}{v^2}\hat\Lambda_{\theta}
\label{Lambda}
\end{equation}

\begin{equation}
\hat\Lambda_{\theta} Y_{l,n}^{(k)} =
\lambda_l Y_{l,n}^{(k)}, \;\;\;
\lambda_l = -l(l+n-2), \;\;\; l= 0,1,2,...,\;\;\;
\label{eigen}
\end{equation}
$$
k= 1,2, 3,....d_{l,n}, \;\;\;\;
d_{l,n}= \frac{(2l+n-2)(n+l-3)}{(n-2)!\;l!}.
$$
The eigenfunctions $Y_{l,n}^{(k)}(\theta)$
are the spherical (harmonic)
functions on $S_1$
 which constitute the orthonormal basis
in $L_2(S_1)$, i.e for any function g({\bf v}) we have:
$$
g({\bf v}) \equiv g(v,\theta) =
\sum_{l=0}^{\infty}\sum_{k=1}^{d_{l,n}} g_{l,k}(v)
Y_{l,n}^{(k)}(\theta)
$$
\begin{equation}
g_{l,k}(v)=\left(Y_{l,n}^{(k)} , g\right)=
\int\limits_{S_1} \overline{Y_{l,n}^{(k)}(\theta)} g(v,\theta){\rm d}S_1 ,
\label{gs1}
\end{equation}
where
\begin{equation}
{\bf v}= (|v|,\theta_1,\theta_2\cdots\theta_{n-1})=(v,\theta),\;\;\;
\theta_1,\cdots\theta_{n-2} \in[0,\pi],\;\;\;
\theta_{n-1} \in [0,2\pi] ,
\end{equation}
\begin{eqnarray*}
v_1&=&v \cos\theta_1, \\
v_2&=&v \sin\theta_1 \cos\theta_2,\\
\cdots&  \cdot&\cdots   \\
v_{n-1}&=&v \sin\theta_1 \cdots\sin\theta_{n-2}\cos\theta_{n-1},\\
v_n&=&v \sin\theta_1 \cdots \sin\theta_{n-2}\sin\theta_{n-1},
\end{eqnarray*}
$$
{\rm d}S_1 = \sin^{n-2}\theta_1 \sin^{n-1}\theta_2\cdots
\sin\theta_{n-2}\,{\rm d}\theta_1{\rm d}\theta_2\cdots{\rm d}\theta_{n-1}.
$$
 Thus for the exact solution of the problem (\ref{P})
we obtain from the Eq.(\ref{Form})

\begin{equation}
P(\lambda,{\bf v})
=\frac{{\rm e}^{\beta_0(\lambda)}}
{\sqrt{4\pi\gamma(\lambda)}}
{\rm e}^{\gamma(\lambda)\hat\Lambda_{\theta}}
\int_{-\infty}^{\infty}
{\rm e}^{-\frac{s^2}{4\gamma(\lambda)}} \varphi\left({\bf v}{\rm e}^
{\beta_1(\lambda)+(n-2)\gamma(\lambda)\pm s}\right){\rm d}s
\label{Psol}
\end{equation}
$$
=\frac{{\rm e}^{\beta_0(\lambda)}}
{\sqrt{4\pi\gamma(\lambda)}}\int_{-\infty}^{\infty}
{\rm e}^{-\frac{s^2}{4\gamma(\lambda)}}
\left[\sum_{l=0}^{\infty}\sum_{k=1}^{d_{l,n}}
{\rm e}^{-\gamma(\lambda)\lambda_l}\,Y_{l,n}^{(k)}(\theta)\,
\varphi_{l,k}\left({\bf v}{\rm e}^
{\beta_1(\lambda)+(n-2)\gamma(\lambda)\pm s}\right)\right]{\rm d}s,
$$
where ${\bf v}\in{\cal R}^n$ , $\beta_0(\lambda), \beta_1(\lambda)$ and
$\gamma(\lambda)$ are from (\ref{beta}),
and $\lambda_l $ and $ d_{l,n} $ are from (\ref{eigen}).

In partucular, for ${\bf v}\in {\cal R}^3$ , we have
($\theta_1\equiv \theta $ and $\theta_2\equiv \varphi$)
$$
\left( \hat\Lambda_{\theta\varphi}=
\frac{1}{\sin\theta}\frac{\partial}{\partial\theta}
\sin\theta\frac{\partial}{\partial\theta} +
\frac{1}{\sin^2\theta}\frac{\partial^2}{\partial\varphi^2}
\right)
$$
\begin{equation}
P(\lambda,{\bf v})
=\frac{{\rm e}^{\beta_0(\lambda)}}
{\sqrt{4\pi\gamma(\lambda)}}
{\rm e}^{\gamma(\lambda)\hat\Lambda_{\theta}}
\int_{-\infty}^{\infty}
{\rm e}^{-\frac{s^2}{4\gamma(\lambda)}} \varphi\left({\bf v}{\rm e}^
{\beta_1(\lambda)+\gamma(\lambda)\pm s}\right){\rm d}s
\label{Psol3}
\end{equation}
$$
=\frac{{\rm e}^{\beta_0(\lambda)}}
{\sqrt{4\pi\gamma(\lambda)}}\int_{-\infty}^{\infty}
{\rm e}^{-\frac{s^2}{4\gamma(\lambda)}}
\left[\sum_{l=0}^{\infty}\sum_{m=-l}^{l}
{\rm e}^{-\gamma(\lambda)l(l+1)}\,Y_{l}^{m}(\theta,\varphi)\,
\varphi_{l,m}\left({\bf v}{\rm e}^
{\beta_1(\lambda)+\gamma(\lambda)\pm s}\right)\right]{\rm d}s ,
$$
where $Y_l^m(\theta,\varphi)$ are the Laplace functions on the 1-sphere
in ${\cal R} ^3$ .

Substituting the expression~(\ref{Psol})
in the Eq.~(\ref{P}) 
one can see immediately that $P(\lambda,{\bf v})$ is a solution of the
problem (\ref{P}) and, according to the Cauchy theorem,
it is the only classical solution of this problem.

\section{Exact solution of the problem (\ref{PP})}
Proceeding as in Section 2, we have for the solution of the problem (\ref{PP}):
\begin{equation}
P(\lambda,{\bf v})={\rm e}^{\beta_0(\lambda)}
{\rm e}^{\beta_1(\lambda){\bf v}\cdot\nabla_{\bf v} +
\gamma(\lambda)({\bf v}\cdot\nabla_{\bf v})^2}\varphi({\bf v}),
\label{Form2}
\end{equation}
where $\beta_0(\lambda),\beta_2(\lambda)$ and $\gamma(\lambda)$
are from (\ref{beta}).

Since
\begin{equation}
\left[\; \beta_1(\lambda){\bf v}\cdot\nabla_{\bf v}\;,\;
\gamma(\lambda)({\bf v}\cdot\nabla_{\bf v})^2\;\right] = 0 ,
\label{com2}
\end{equation}
we can factorize in the Eq.~(\ref{Form2}):
\begin{equation}
P(\lambda,{\bf v})=
{\rm e}^{\beta_0(\lambda)}
{\rm e}^{\beta_1(\lambda){\bf v}\cdot\nabla_{\bf v}}
{\rm e}^{\gamma(\lambda)({\bf v}\cdot\nabla_{\bf v})^2}
\varphi({\bf v}) =
{\rm e}^{\beta_0(\lambda)}
{\rm e}^{\gamma(\lambda)({\bf v}\cdot\nabla_{\bf v})^2}
{\rm e}^{\beta_1(\lambda){\bf v}\cdot\nabla_{\bf v}}
\varphi({\bf v}).
\label{Form3}
\end{equation}
Now using (\ref{F1}) and the formula
\begin{equation}
{\rm e}^{\gamma(\lambda)({\bf v}\cdot\nabla_{\bf v})^2}
\varphi({\bf v}) =
\frac{1}
{\sqrt{4\pi\gamma(\lambda)}}
\int_{-\infty}^{\infty}
{\rm e}^{-\frac{s^2}{4\gamma(\lambda)}} \varphi\left({\bf v}{\rm e}^{\pm s}
\right){\rm d}s  ,
\label{F3}
\end{equation}
we find for the exact solution  of the problem (\ref{PP}):
\begin{equation}
P(\lambda,{\bf v}) =
\frac{{\rm e}^{\beta_0(\lambda)}}
{\sqrt{4\pi\gamma(\lambda)}}
\int_{-\infty}^{\infty}
{\rm e}^{-\frac{s^2}{4\gamma(\lambda)}} \varphi\left({\bf v}{\rm e}^
{\beta_1(\lambda)\pm s}
\right){\rm d}s,
\label{Psol2}
\end{equation}
where $\beta_0(\lambda),\beta_1(\lambda)$ and $\gamma(\lambda)$
are defined in~(\ref{beta}).

Substituting the expression (\ref{Psol2}) in the Eq.(\ref{PP})
one can check that $P(\lambda,{\bf v})$
 is solution of the problem (\ref{PP}),
and, according to the Cauchy theorem, it is the only classical
solution of the problem.
\section{Conclusions}

The exact solutions of the Cauchy problems (\ref{P}) and (\ref{PP})
are obtained, using the disentangling techniques of R. Feynman and
M. Suzuki's methods for solving the
\mbox{1-dimensional} Fokker-Planck
equation. The problems (\ref{P}) and (\ref{PP}) are two of many
n-dimensional generalizations of the 1-dimensional FPE, used by
Friedrich and Peinke in their description of a turbulent cascade
as a stochastic process of marcovian type.

\end{document}